\pdfoutput=1
\documentclass{JINST}
\usepackage{lineno}

\title{High Resolution Gamma Ray Detection in a Dual Phase Xenon Time Projection Chamber}

\author{Q.~Lin$^a$, Y.~Wei$^a$, J. Bao$^c$, J.~Hu$^a$, X. Li$^d$, W.~Lorenzon$^b$, K.~Ni$^a$\thanks{Corresponding author.}, M.~Schubnell$^b$, M.~Xiao$^a$, X.~Xiao$^a$\\
\llap{$^a$}  INPAC, Department of Physics \& Astronomy and Shanghai Key Laboratory for Particle Physics and Cosmology, 
Shanghai Jiao Tong University, 800 Dongchuan Road, Shanghai, 200240, P. R. China\\
\llap{$^b$}  Randall Laboratory of Physics, University of Michigan, Ann Arbor, Michigan 48109-1040, USA \\
\llap{$^c$}  Science and Technology on Nuclear Data Laboratory, China Institute of Atomic Energy, \\
Beijing, 102413, P. R. China\\
\llap{$^d$}  School of Physics, Peking University, Beijing, 100871, P. R. China\\
  E-mail: \email{nikx@sjtu.edu.cn}
}

\abstract{
Dual phase Xenon Time Projection Chambers (XeTPCs) are being used by several experiments as a promising technique for direct detection of dark matter.  We report on the design and performance of a small 3-D sensitive dual phase XeTPC. The position resolution is 2\,mm in the center of detector, limited by the hole size of the mesh at the proportional scintillation region. An energy resolution of $1.6\%$($\sigma/E$) for 662\,keV gamma rays is achieved in the very center of the detector by combining the ionization and scintillation signals at a drift field of 0.5\,kV/cm. This represents the best energy resolution achieved among liquid xenon detectors to date. The energy resolution is only slightly dependent on the drift field. Better than 2\% energy resolution ($\sigma/E$) for 662\,keV gamma rays was achieved for drift fields between 100\,V/cm and 2\,kV/cm. With high position and energy resolution, a dual phase XeTPC has also potential applications in surveys 
for 
neutrinoless double-beta decay and in gamma ray imaging. 
}


\begin{document}


\section{\label{sec:introduction} Introduction}
Xenon Time Projection Chambers (XeTPCs) are widely used in particle physics and astrophysics applications, including dark matter searches\,\cite{XENON10, XENON100, ZEPLINII, ZEPLINIII, LUX}, neutrinoless double-beta decay surveys\,\cite{EXO-200, EXO-PRL}, gamma-ray astrophysics\,\cite{ComptonTelescope1, ComptonTelescope2, ComptonTelescope3} and medical imaging\,\cite{MedicalImaging}. Recently developed XeTPCs, such as those used in the XENON100 and EXO experiments, have achieved high 3-D position resolutions (typically 3\,mm) and good energy resolutions ($\sigma/E<5\%$ above 1\,MeV).

A dual phase (liquid/gas) XeTPC detects the ionization signals from liquid xenon by drifting the electrons into the gas phase, where the electrons are amplified in a strong field ($\sim$ 10\,kV/cm) to produce electroluminescence, or so called proportional scintillation light (S2)\,\cite{Aprile:IEEE04}. The signal is typically measured by low noise photomultiplier tubes (PMTs). The primary scintillation signals (S1) in the liquid xenon are detected by PMTs as well.  This technique allows the detection of very low energy recoils (a few keV) and has the capability to discriminate between nuclear and electron recoils based on the ratio between S1 and S2\,\cite{Aprile:PRL06}. Combined with 3-D position reconstruction capability and event-type discrimination for background reduction, the LUX dual phase XeTPC has reached the best sensitivity for spin-independent elastic interactions of weakly interacting massive particles (WIMPs) with nucleons\,\cite{LUX}. 

A dual phase XeTPC has the advantage of being able to detect very low energy events with lower electronic noise than charge readout using a pre-amplifier. Even a single electron drifting from the liquid xenon can be detected\,\cite{single-e}. Thus a dual phase XeTPC has the potential to reach a better energy resolution than a conventional liquid XeTPC using charge readout, provided that the variance in the amplification gain itself is small. This makes the dual phase XeTPC attractive for gamma-ray imaging and neutrinoless double beta decay surveys where high energy resolution is crucial. 

Low variance of the amplification gain and high energy resolution can be achieved in a TPC by removing the position-dependence of the S1 and S2 signals through precise reconstruction of the 3-D position of each event.
The position dependence of the S1 and S2 signals are mainly introduced by the variation of the light collection efficiency over the sensitive volume and the proportional light region. Liquid levels that are not uniform
can also cause an X-Y dependence of the S2 signals.

Due to the common requirement of ultra-low radioactive background for dark matter and neutrinoless double beta decay surveys using liquid xenon, it is possible to search for two types of signals simultaneously using a dual phase XeTPC~\cite{Arisaka:XAX}. While a dark matter detector is optimized to detect low energy events in the few keV to tens of keV energy range, a double beta decay detector is looking for signals at around 2.5\,MeV. Such combined searches require modification of signal readout from single-purpose searches.

In this paper, the design and setup of the liquid xenon detector equipped with a small XeTPC are introduced in Sec.~\ref{sec:setup}. This is followed by demonstrating methods to obtain the 3-D positions and their application to correct the position dependence of the signals. The results and discussion of the energy resolution at different drift fields for energy gamma rays of various energies in the small XeTPC are illustrated in Sec.~\ref{sec:results}.

\section{\label{sec:setup} Detector design and setup} 
\begin{figure}
 \centering
 \includegraphics[height=7.5cm]{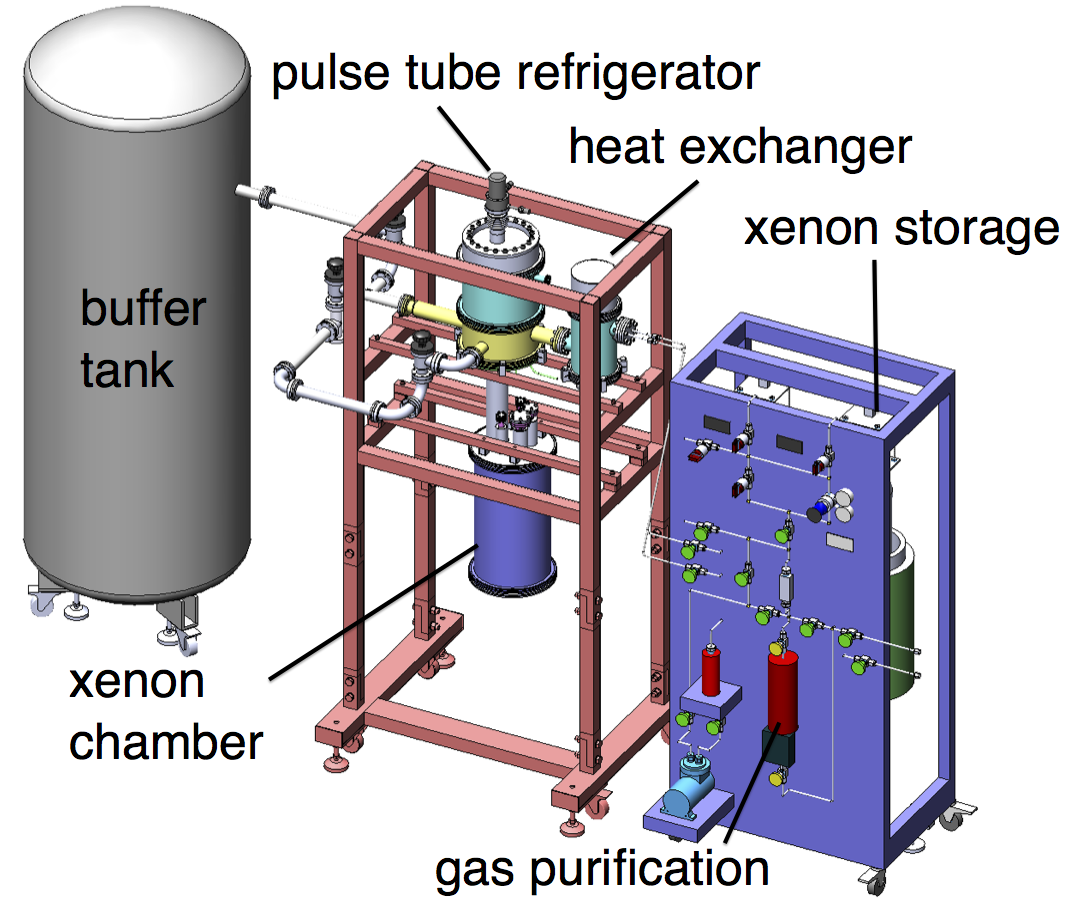}
 \caption{\small Schematic view of the liquid xenon cryogenic and purification system. Shown are the pulse tube refrigerator, the heat exchanger and the vacuum-insulated chamber for liquid xenon. A gas purification system is used to purify the gas continuously during the operation. A buffer tank allows to automatically recover xenon gas during an emergency.}
 \label{fig:CryogenicsDrawing}
\end{figure}

For stable operation of the liquid xenon detector and efficient cooling and purification of the xenon gas, a cryogenic system (Fig.~\ref{fig:CryogenicsDrawing}) consisting of a pulse tube refrigerator (PTR) from Iwatani corporation, a heat exchanger and a vacuum insulated xenon chamber was built. The PTR consists of a PDC08 cold head and an SA115 helium compressor, providing 40\,W of cooling power at 165\,K, the temperature of liquid xenon. During operation liquefied xenon drips through a 1/4$^{\prime\prime}$ stainless steel tube into the inner xenon chamber where the small XeTPC is located. 

In total, about 10\,kg of liquid xenon is required to fill the chamber for operation. Xenon gas is stored in two high pressure aluminum bottles, each with a volume of 8\,liters and capable of storing up to 10\,kg of xenon gas.  The two bottles are connected to a gas purification system as shown in Fig.~\ref{fig:CryogenicsDrawing}. A SAES purifier (model PS3-MT3-R-2) is used to remove impurities such as water and oxygen from xenon gas down to part-per-billion (ppb) levels. Continuous circulation and purification of the xenon gas at a flow rate of at least 5\,SLPM (standard liter per minute) during detector operation is achieved by using a diaphragm circulation pump (Enomoto MX-808ST-S) and a flow controller (Teledyne model HFC-302). A heat exchanger, placed at the inlet and outlet of the detector, saves cooling power during gas circulation\,\cite{Giboni:heatex}. 

A slow control system was implemented for recording operational data and monitoring detector performance through some crucial parameters, such as the temperature and pressure in the detector. The slow control system is connected to UPS battery power. Alarm messages are sent to cell phones in case of an emergency situation. To recover xenon, liquid nitrogen is utilized to transfer the gaseous xenon from the detector to an
aluminum bottle. The detector is also connected through a battery-operated solenoid valve to a buffer tank that is rated up to 4\,atm. During an emergency when liquid nitrogen is not available, the buffer tank can temporarily store up to 16\,kg of xenon. The system ran in stable condition during the entire experimental period ($\sim$ 50 days) reported here.

\begin{figure}[htp]
 \centering
 \includegraphics[height=9cm]{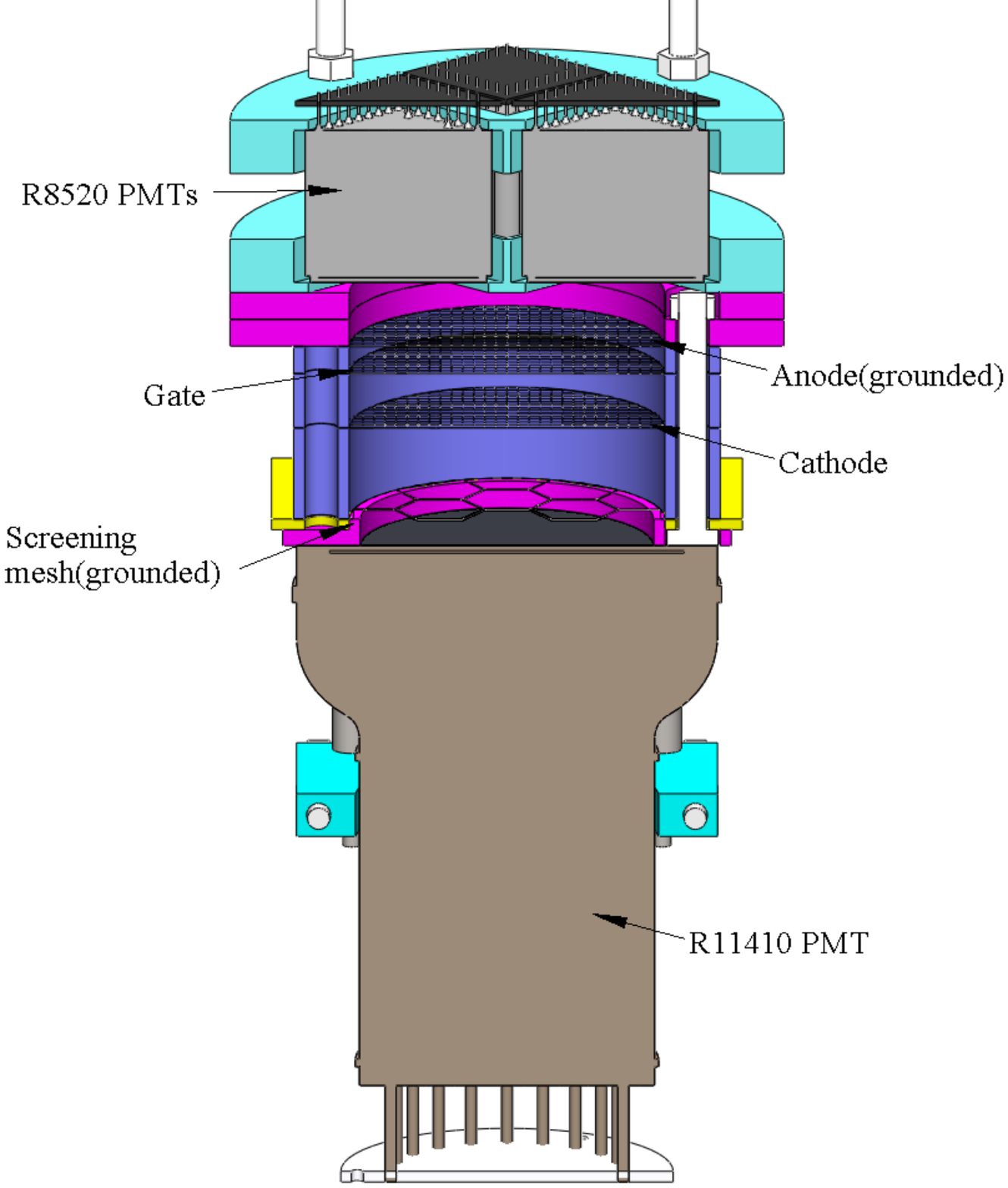}
  \caption{\small Schematic cross section of the time projection chamber. Three etched meshes define the drift and gas scintillation region. Four R8520 PMTs on the top and one R11410 PMT on the bottom are used. A hexagonal mesh with high transparency is used as a shielding mesh to protect bottom PMT.}
  \label{fig:TPCschematic}
\end{figure}

A small XeTPC structure, as shown in Fig.~\ref{fig:TPCschematic}, is attached to the top flange of the liquid xenon vessel by PEEK
rods. The TPC has three electrodes which are denoted as anode, gate (or grid) and cathode. Etched stainless steel
mesh was used for the electrodes. The mesh pitch is 2\,mm and the stainless steel thread diameter is 0.1\,mm, resulting in an optical transparency of 92\%. The distance between the anode and gate is 5\,mm. The liquid level is adjusted by regulating the circulation flow rate to provide a proper gas
gap for the proportional scintillation light. 
A liquid level meter is used to monitor the fluctuation of liquid level. During a typical data-taking period of one day, the level fluctuation was measured to be 0.11\,mm (1$\sigma$ variantion). The absolute 
gas gap was measured by using the small S2 signals generated by photoionization electrons from the cathode electrode according to a method described in\,\cite{Xenon100_SingleE}.
Typical gas gap values of 2.6$\pm$0.5\,mm for the activated xenon measurement and 1.4$\pm$0.4\,mm for the $^{137}$Cs measurement were obtained.
The distance between the gate and cathode is 1\,cm, defining the total drift volume, or
sensitive volume for the gamma ray detection. A hexagonal screening mesh with a pitch of 14\,mm and a thread diameter of 0.2\,mm, giving an optical transparency of 97.5\%, 
is placed 5\,mm above the bottom PMT and connected to ground. This screening electrode
is located about 1.7\,cm below the cathode. The electrodes are sandwiched between Teflon blocks with inner diameters of 57\,mm, defining the diameter of the
cylindrical sensitive volume.
 
\begin{figure}[htp]
 \begin{minipage}{0.5\textwidth}
  \centering
  \includegraphics[height=6cm, width=7.5cm]{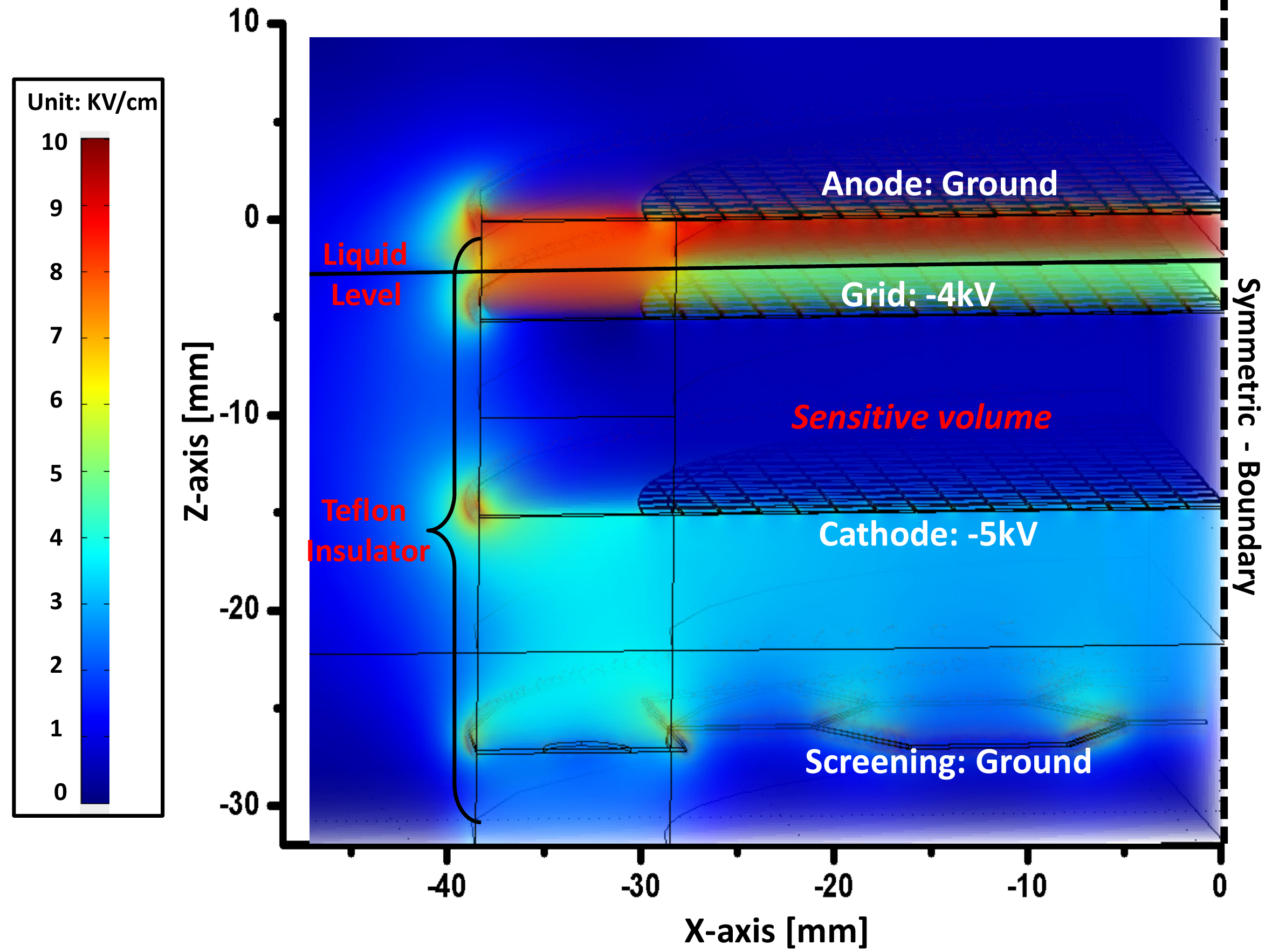}
 \end{minipage}
\hfill
 \begin{minipage}{0.5\textwidth}
 \centering
  \includegraphics[height=6cm, width=7.5cm]{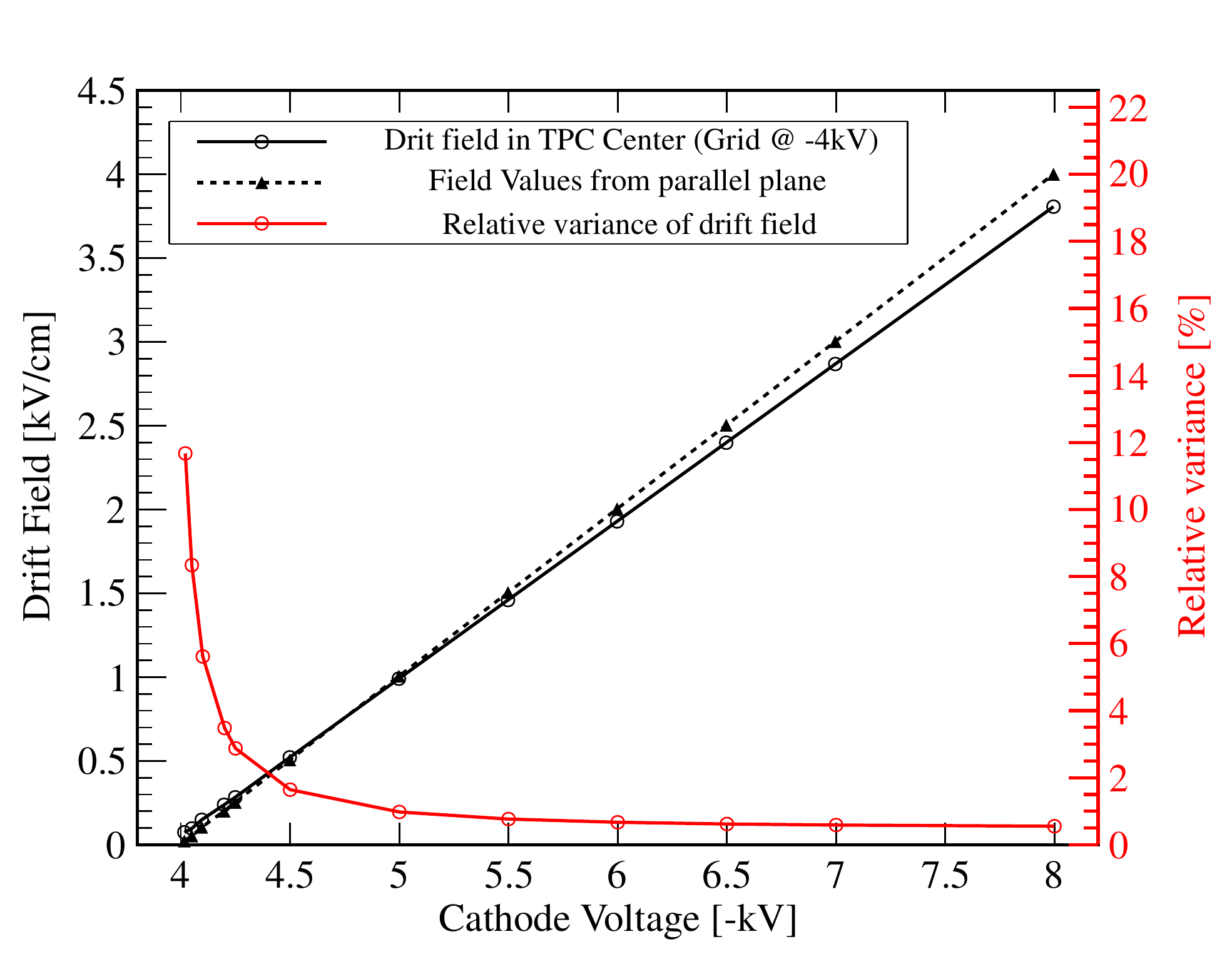}
 \end{minipage}
 \caption{\small (Left) A 3-D field simulation using a realistic detector geometry, including the exact shape and dimension of the mesh electrodes, using the COMSOL software. (Right) Simulated results for the mean and relative variance of drift field in the sensitive volume.}
 \label{fig:tpcfield}
\end{figure}

During operation, the anode electrode is connected to ground. Negative voltages are supplied to the gate and the cathode to provide appropriate electric fields to drift electrons and amplify them in the gas gap. An electric field simulation was performed to investigate the field uniformity in the TPC using the finite element analysis software package COMSOL\,\cite{COMSOL}. Fig.~\ref{fig:tpcfield} shows a typical configuration with the grid at -4\,kV, the cathode at -5\,kV, and both the anode and the screening meshes grounded. According to the simulation, a very small deviation from the expected field from two infinitely large parallel plates is obtained. The absolute deviation is less than 75\,V/cm when the cathode voltage is above -6\,kV, which corresponds to a drift field of 2\,kV/cm, and reaches 200\,V/cm when the cathode voltage is -8\,kV (4\,kV/cm drift field). The relative variance of the field across the drift volume decreases 
with increasing cathode voltage (absolute value) 
and drops to below 1\% when the cathode voltage is below -5\,kV.

PMTs are placed at the top and the bottom of the TPC structure to detect the scintillation light. The top PMT array, placed 10\,mm above the anode electrode, consists of four Hamamatsu R8520 PMTs. Each covers an optical area of about $1^{\prime\prime}\times1^{\prime\prime}$. The distribution of the proportional scintillation light (S2) on the four top PMTs can be used to reconstruct the X-Y position of an event. One Hamamatsu R11410 PMT with an outer diameter of about 3$^{\prime\prime}$ is placed at the bottom. The R11410 PMT\,\cite{3-inchPMT} has a better single-photoelectron resolution and a better photoelectron collection efficiency to the first dynode than the R8520 PMTs. This makes it a preferable bottom PMT because most primary scintillation light (S1) is reflected at the liquid surface, leading to a much higher light intensity at the bottom than at the top. 

\begin{figure}[htp]
 \centering
 \includegraphics[height=9cm]{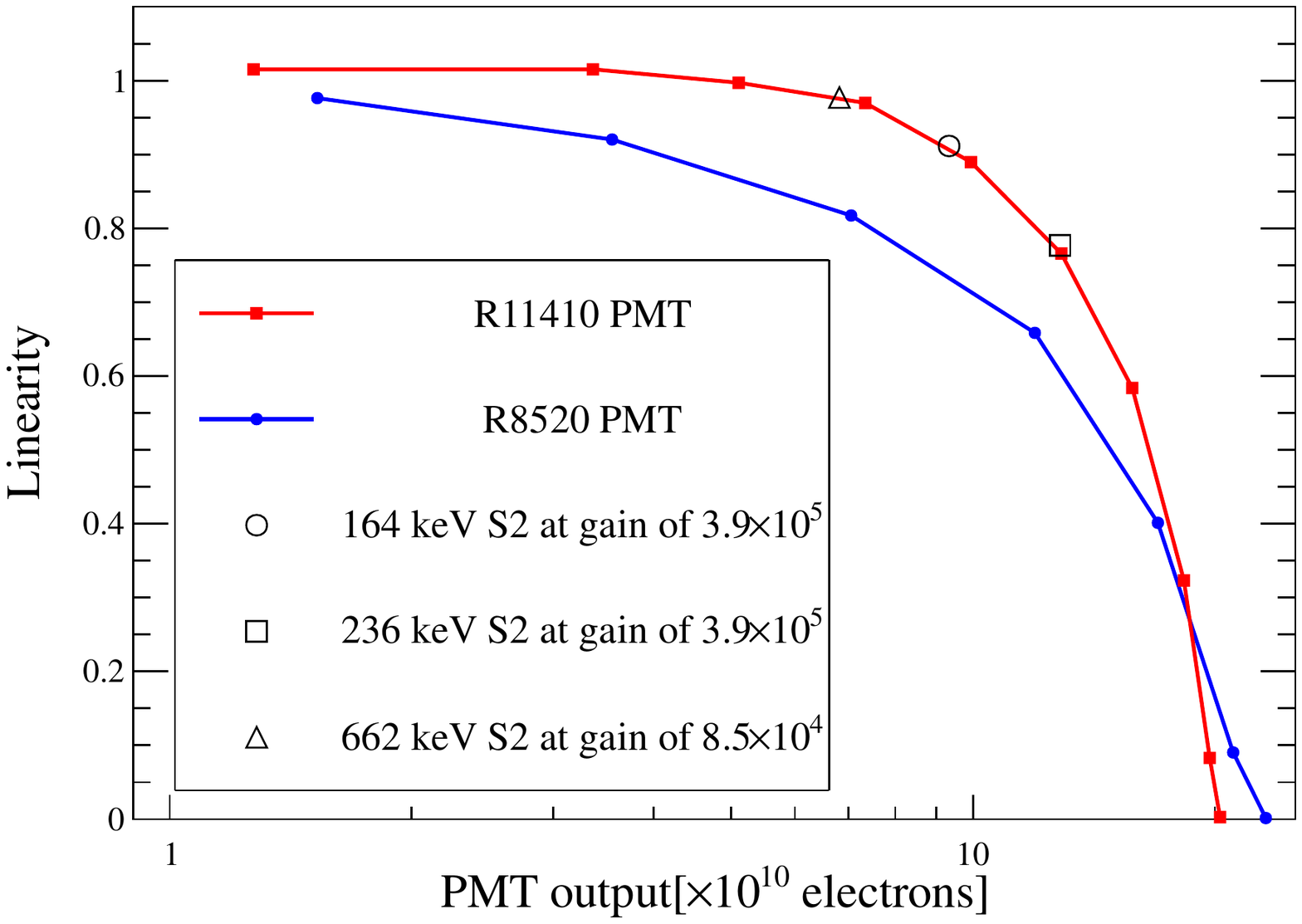}
  \caption{\small Measured linearity of R8520 and R11410 PMTs. The R11410 PMT displays a linearity of better than 0.8 for signals below 1.2$\times$10$^{11}$ electrons. It saturates when the signal reaches 2.0$\times$10$^{11}$ electrons. The R8520 PMT has a linearity of 0.8 once the signal reaches 7.5$\times$10$^{10}$ electrons, and saturates with 2.2$\times$10$^{11}$ electrons. Open symbols indicate the linearity of the bottom PMT responses to S2 signals from 164, 236 and 662\,keV gamma rays. }
  \label{fig:Linearity}
\end{figure}

\begin{figure}[htp]
 \centering
\includegraphics[width=11cm]{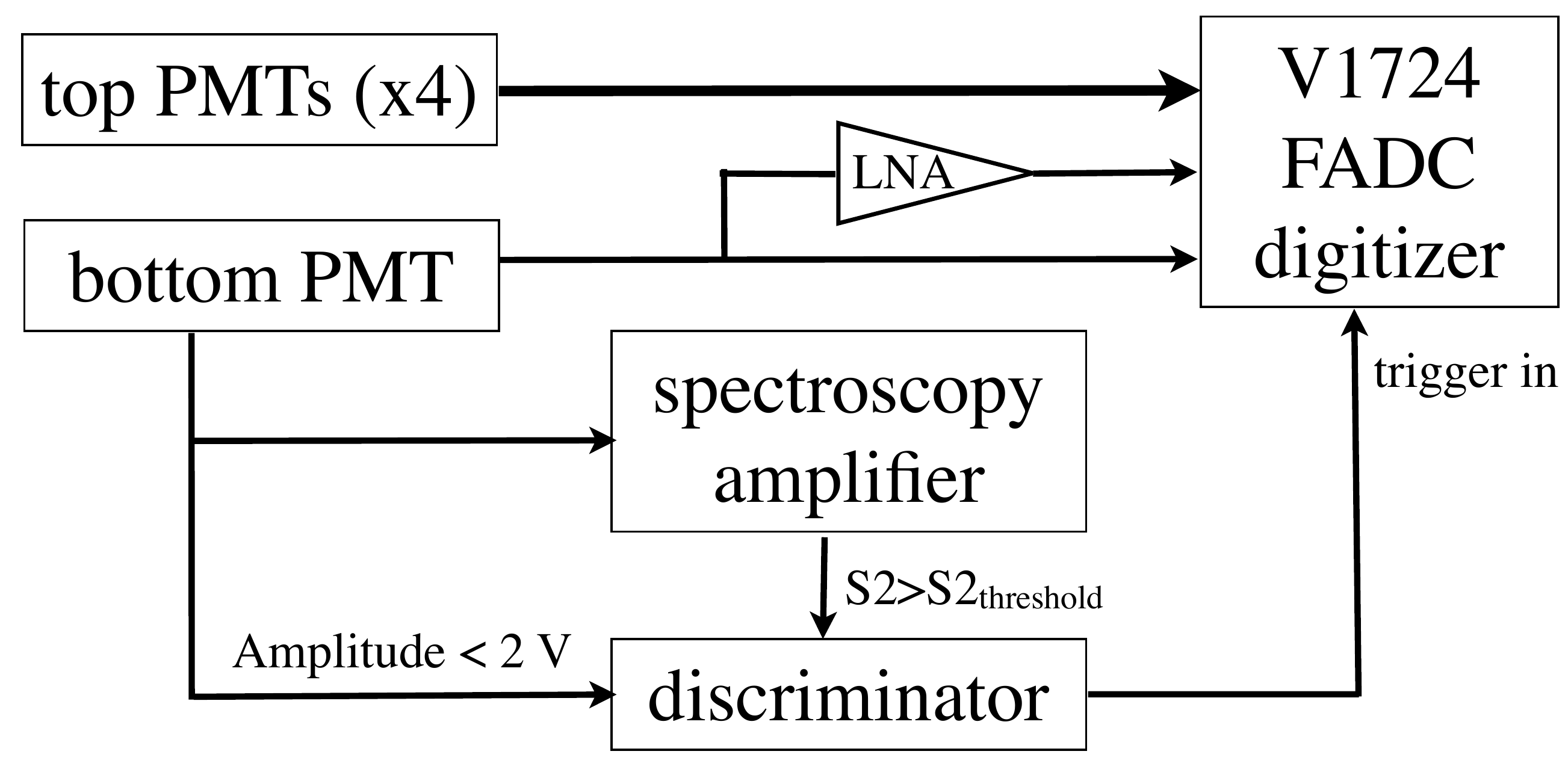}
\includegraphics[width=11cm]{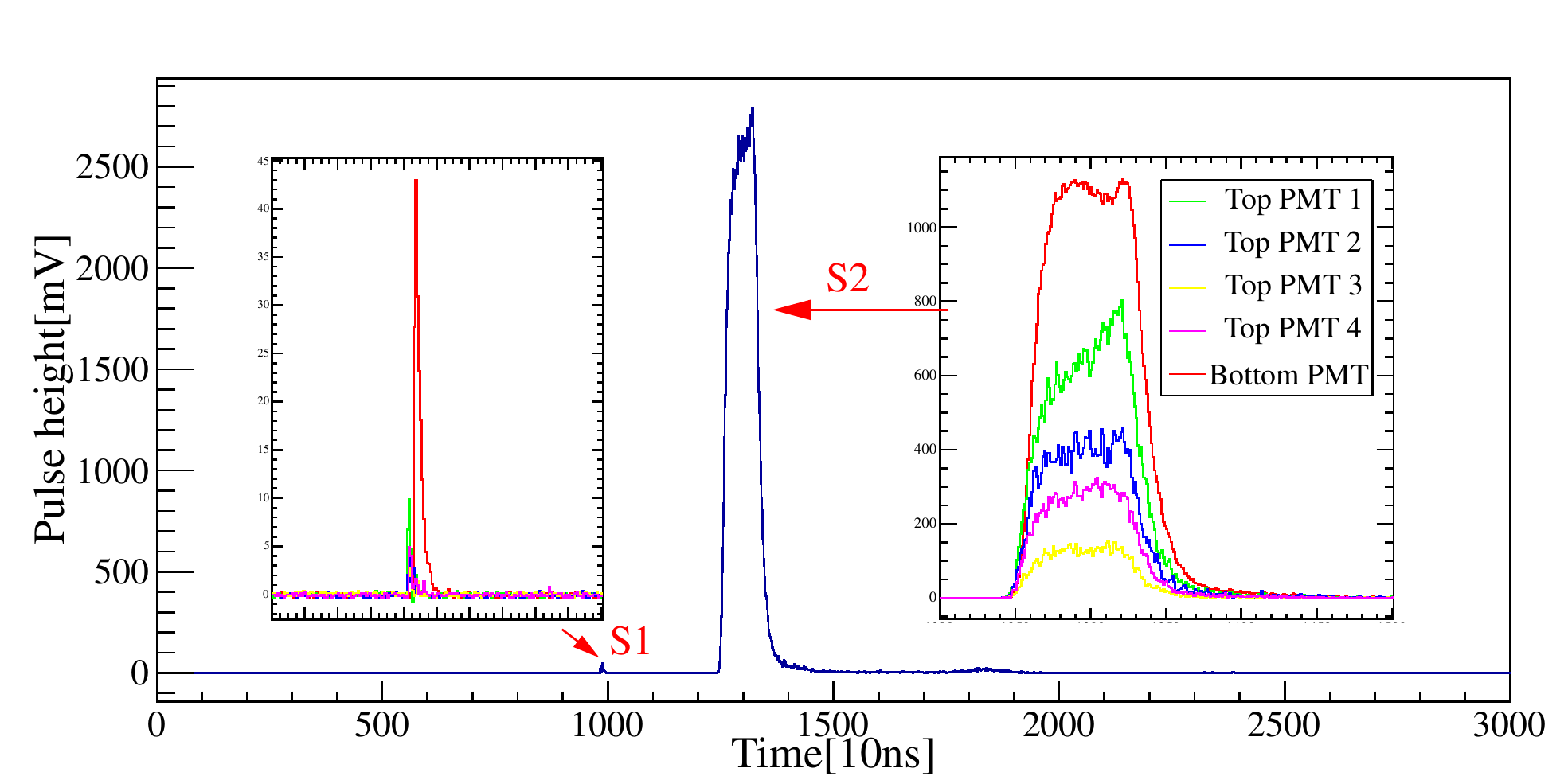}
  \caption{\small Top: Schematic diagram showing the data readout electronics (See text for details). Bottom: A typical event waveform, from a 164\,keV gamma ray, summed over all PMTs. The insets show the S1 and S2 signal from each individual PMT. The total S1 has 625\,PE and a S2 response of 273,748\,PE is observed by the bottom PMT. }
  \label{fig:daq}
\end{figure}

Two radioactive sources were used for measurements: neutron-activated xenon and $^{137}$Cs. 
The PMT gains were calibrated both before the measurements of radioactive sources in cold xenon gas and after the measurements of radioactive 
sources in liquid xenon temperature. The gains calibrated in liquid xenon temperature were used in our analysis.
The activated xenon data were taken with the top PMTs operated at gains of about 10$^6$ and the bottom PMT at a gain of 3.9$\times$10$^5$. $^{137}$Cs
data were taken with the top and bottom PMTs at gains of about 5$\times$10$^5$ and 8.5$\times$10$^4$, respectively. The gain fluctuation is less than 2\% within two weeks.
The bottom PMT was operated at a relatively low gain in comparison with the top PMTs, in order to avoid PMT
saturation for high energy events. The linearity of the PMT response (Fig.~\ref{fig:Linearity}) 
at liquid xenon temperature was measured with two synchronized LEDs with one 
LED (LED$_1$) at a fixed light output and the other one (LED$_2$) with varying intensity. We define the linearity
according to
\begin{equation}
L = \frac{ S_{LED_1+LED_2} - S_{LED_2} }{ S_{LED_1} } ,  
\label{equ:Linearity}
\end{equation}
where $S_{LED_1}$, $S_{LED_2}$ and $S_{LED_1+LED_2}$ represent the measured signal intensity for LED$_1$ only, LED$_2$
only and both LEDs. From the data from Fig.~\ref{fig:Linearity} and the S2 values obtained from measurements with
radioactive sources we estimate a linearity of at least 0.91, 0.77 and 0.98 for gamma rays at energies of 164\,keV,
236\,keV and 662\,keV, respectively. The linearity response of the R8520 PMT is worse than that of the R11410 PMT. In the analysis we use only S2
from the bottom R11410 PMT to represent the S2 signal. The PMT waveforms were digitized by a CAEN V1724 FADC with a sampling frequency of 100\,MS/s and a bandwidth of 100\,MHz. Traces with length of $40 \mu s$ with 33\% post trigger were taken and transferred to a computer for offline analysis. The data acquisition 
system, shown in Fig.~\ref{fig:daq}, can be configured to either detect low energy events, such as those related to dark matter searches, or high energy events relevant for gamma ray imaging and neutrinoless double beta decay surveys.\footnote{For the measurements presented here, only the high energy readout configuration was used.}

The bottom PMT's signal was split into two. One was read out directly for measuring high energy signals, while the other was fed into a low noise amplifier (LNA-1440) for measuring low energy events. For high energy events, we required a five-fold coincidence for an event trigger. For triggering on low energy events, the bottom PMT's signal was fed into a spectroscopy amplifier, the output of which was sent to a discriminator. Events with S2 larger than a few hundred photoelectrons were triggered. The bottom PMT's signal was also fed directly to a discriminator to provide a veto on pulse height for high energy events, requiring the pulse amplitude to be less than 2 V. In principle, these two trigger modes can be operated simultaneously without adding dead time to the system.

\section{\label{sec:results} Results and Discussion} 

In order to measure the performance of the position reconstruction, neutron-activated xenon was used to provide a gamma-ray source uniformly distributed in the detector's sensitive volume. A stainless steel sampling bottle containing approximately 1\,kg of natural xenon was activated at the China Institute of Atomic Energy (CIAE) by using a $5\times10^6$\,neutron/s $^{252}$Cf source placed 1\,cm away from the wall of the xenon bottle. This procedure is similar to that described in Ref.\,\cite{Ni:NIM07} but without a polyethylene container surrounding the bottle. It resulted in activated xenon states of $^{129m}$Xe (E$_\gamma$=236.1\,keV, T$_{1/2}$=8.9\,days) and $^{131m}$Xe (E$_\gamma$=163.9\,keV, T$_{1/2}$=11.8\,days) with activity of about 4\,kBq/kg. A $^{137}$Cs source was placed outside of the xenon chamber to verify the response of the detector to 662\,keV gamma rays.

\subsection{\label{subsec:posres} Position resolution}

In a dual phase xenon detector, the 3-D position of an event can be determined. The $z$ position is determined through the drift time of electrons from the event's primary location to the liquid-gas interface with sub-mm resolution\,\cite{XENON100}. The X-Y position is obtained from the S2 signal's distribution on the top PMT array by using position reconstruction algorithms such as Neural Network (NN)\,\cite{NN} and Support Vector Machine (SVM)\,\cite{SVM} through comparison with Monte Carlo data.
The X-Y position resolution in a dual phase TPC is dominated by two contributions: drift electron transport and event X-Y position reconstruction. The drift electron transport contribution ($\sigma_{drift}$) is due to the diffusion of electron clusters during drifting and the curvature of the electric field lines along the electron trajectory. In the setup, the detector has a maximum drift length of only 1\,cm, which corresponds to a maximum drift time of 5\,$\mu$s at 1\,kV/cm. The transverse diffusion coefficient is approximately 100\,cm$^2$/s\,\cite{DT}. This results in a maximum transverse diffusion of 0.22\,mm. The field lines tend to curve around the gate mesh, introducing a contribution of 2\,mm. This contribution is much larger than that from diffusion, and thus dominates the X-Y position resolution.

\begin{figure}[htp]
 \begin{minipage}{0.5\textwidth}
  \centering
  \includegraphics[height=6cm, width=7.5cm]{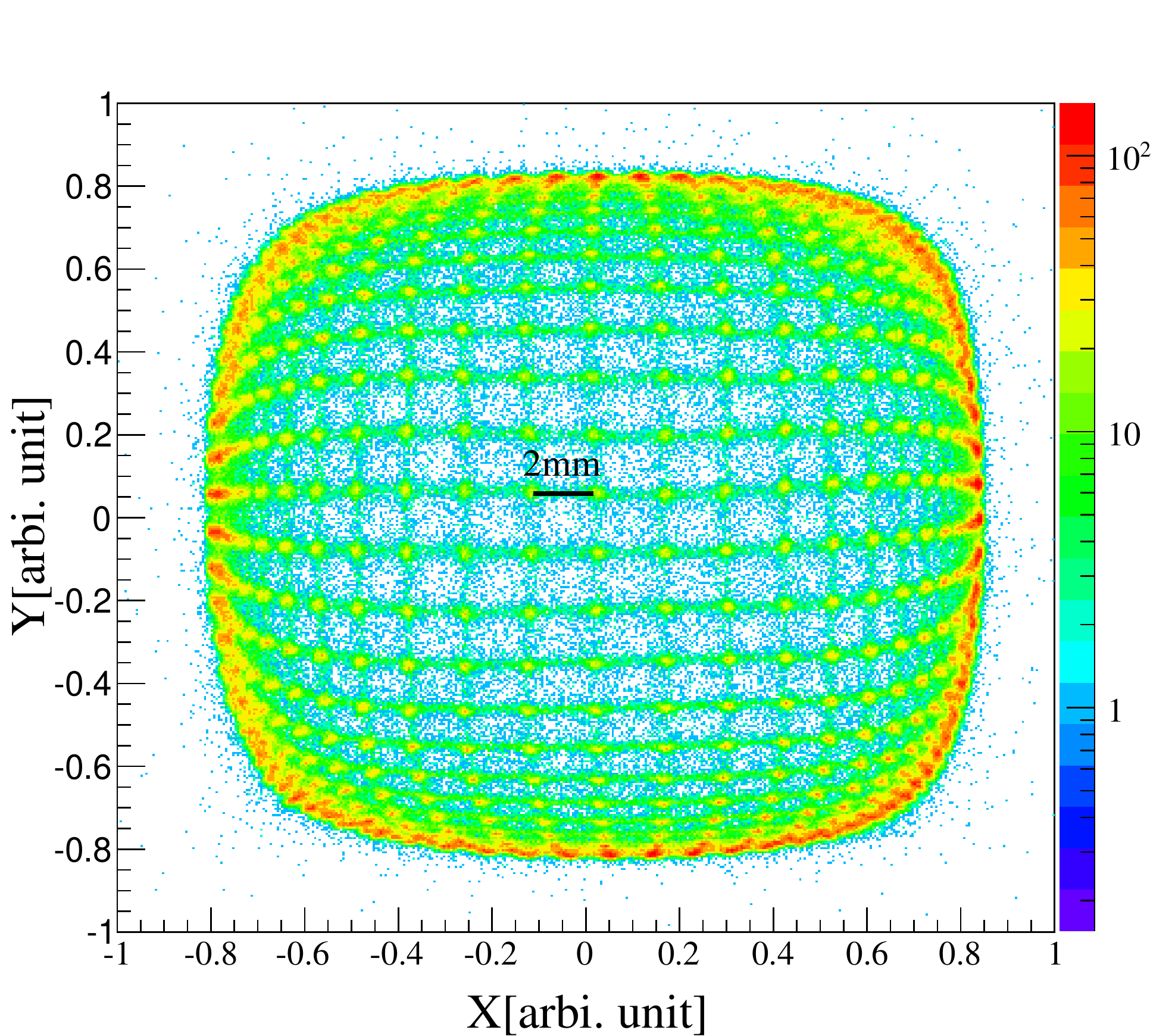}
 \end{minipage}
\hfill
 \begin{minipage}{0.5\textwidth}
 \centering
  \includegraphics[height=6cm, width=7.5cm]{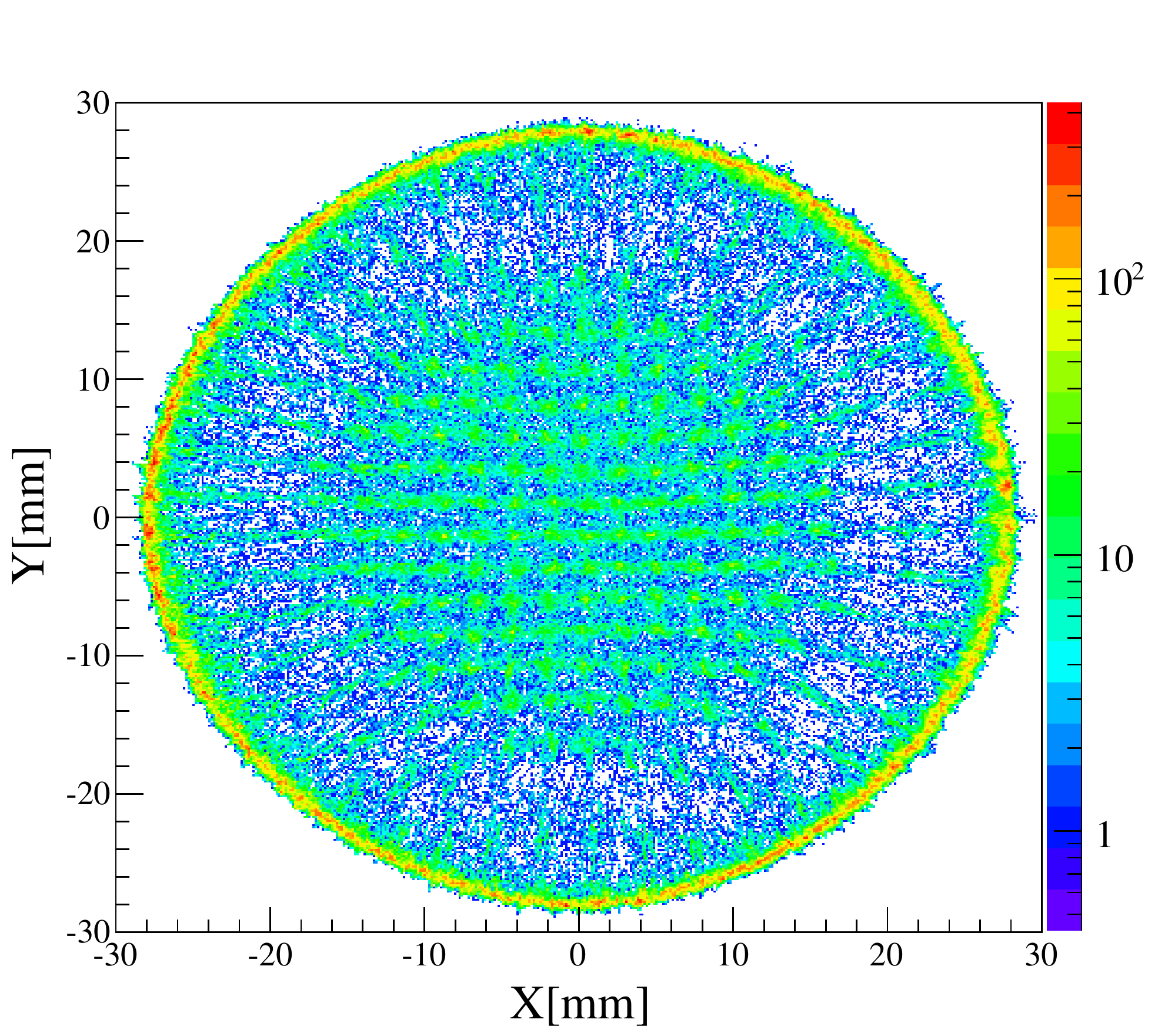}
 \end{minipage}
 \caption{\small (Left) Reconstructed X-Y position based on center-of-gravity for gamma rays from activated xenon. The effect of the anode mesh structure (2\,mm $\times$ 2\,mm) in the central part is clearly visible. In the central region, a 2-D Gaussian function was used to fit the points and a reconstruction variance ($\sigma_{rec}$) of 0.37\,mm is obtained. (Right) The reconstructed X-Y position based on Monte Carlo comparison of S2 pattern on top PMTs from same activated xenon run as in left panel. It shows the S2 pattern is sensitive to events in the central region within a radius of around 15\,mm.}
 \label{fig:xypos}
\end{figure}

The event X-Y position reconstruction contribution ($\sigma_{rec}$) is due to the photon detection devices' position,
arrangement and detection efficiency. To reconstruct the X-Y position, a simple algorithm based on center-of-gravity was used, such that
\begin{equation}
\vec{r} = \frac {\Sigma^4_{i=1} N^i_{pe} \vec{r}_i}{\Sigma^4_{i=1} N^i_{pe}} ,
\label{equ:positionreconstruction}
\end{equation}
where $N^i_{pe}$ is the number of detected photoelectrons for each top PMT, and $\vec{r}_i$ represents the projection vector of each top PMT's center on X-Y. The reconstructed positions, $\vec{r}$, for events from an activated xenon run are shown in the left panel of Fig.~\ref{fig:xypos}. Due to the electric field lines focusing on the anode mesh wires, the S2 light is mostly generated below the wires. A position pattern showing the mesh pitch of 2\,mm$\times$2\,mm is clearly visible. In the central part of the TPC, we can calibrate the reconstructed position by using the pitch (2\,mm) between two adjacent wires. From the calibration and a 2-D Gaussian fitting procedure of the position distribution around the wire crossing point, an X-Y position variance due to event reconstruction ($\sigma_{rec}$) of 0.37$\pm$0.06\,mm in the center of detector is obtained. The uncertainty is calculated from the variance of fitted values 
in 
the 
nine most central points.

Therefore the position resolution ($\sigma_{x,y}$) in the very center of detector is 2\,mm, dominated by the hole
size of gate mesh in comparison with relative small reconstruction contribution (0.37\,mm). Further optimization
by using a finer mesh will improve the X-Y resolution in the center.

The reconstructed position scale is not uniform across X-Y, which is because the top PMTs are insensitive to the S2 pattern for events near the edge. The reconstructed distance between two adjacent crossing points in the center of detector is 2.5 times larger than that at radius of 14\,mm. The same reason also caused the larger reconstructed position resolution near the edge. We also developed a position reconstruction method based on comparison of data with Monte Carlo simulated S2 pattern on the top PMTs. The right panel of Fig.~\ref{fig:xypos} shows the reconstructed position based on this method from the same activated xenon run as shown in the left panel. The result also shows a non-uniform position scale, which confirms that the non-uniformity comes from the non-sensitivity of the top PMTs to S2 pattern for edge events. In our analysis, we used the reconstructed position based on center-of-gravity for the fiducial volume cut and position-dependent signal correction.

\subsection{\label{subsec:Signal Correction} Detector performance and signal correction}
\begin{figure}[htp]
 \centering
 \includegraphics[height=7cm, width=11cm]{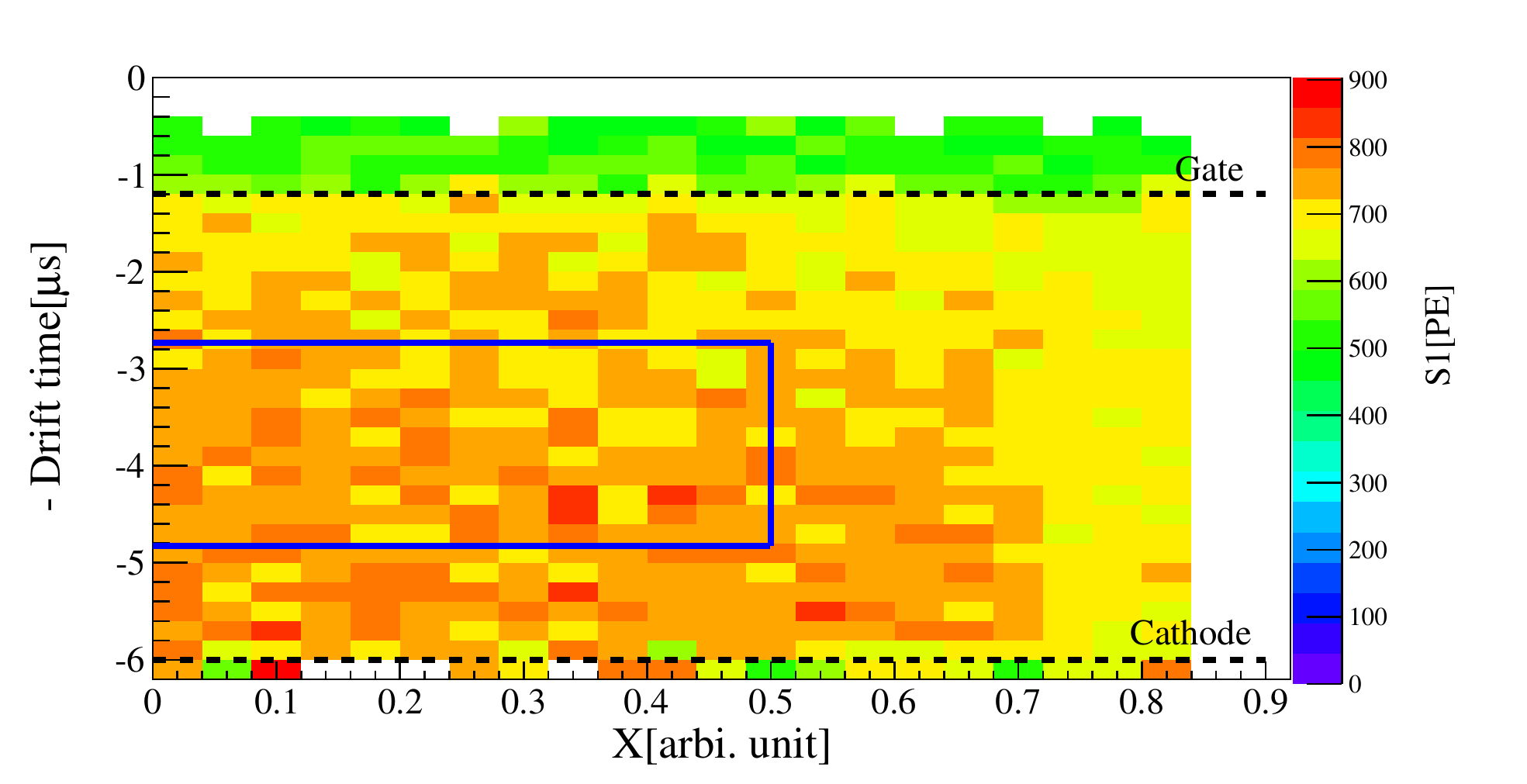}
 \caption{\small The position dependence of the S1 signal in the selected volume (|Y|<0.25) for 164\,keV gamma rays for a drift time of 1\,kV/cm. The blue solid line indicates the fiducial volume selection of central events. This corresponds to a cylinder with a height of 4.2\,mm and diameter of 18\,mm, while dashed lines indicate the gate and cathode positions.}
 \label{fig:S1posdep}
\end{figure}

Both the S1 and S2 signal amplitude depend on the event location because of the non-uniform light generation caused by drift field variations, and of the non-uniform light collection at the different locations of the sensitive volume. The good 3-D position resolution and the uniformly distributed activated xenon source allowed us to determine the position dependence of the two signals. Fig.~\ref{fig:S1posdep} shows the S1 position dependence for 164\,keV gamma rays in the TPC at a drift field of 1\,kV/cm. The S1 signal variance (RMS/mean) in the entire region is 6.0\%. After correcting the position dependence using the map from Fig.~\ref{fig:S1posdep}, the over position variance decreases to 4.2\%.

\begin{figure}[htp]
 \begin{minipage}{0.48\textwidth}
  \centering
  \includegraphics[height=5.5cm, width=7.5cm]{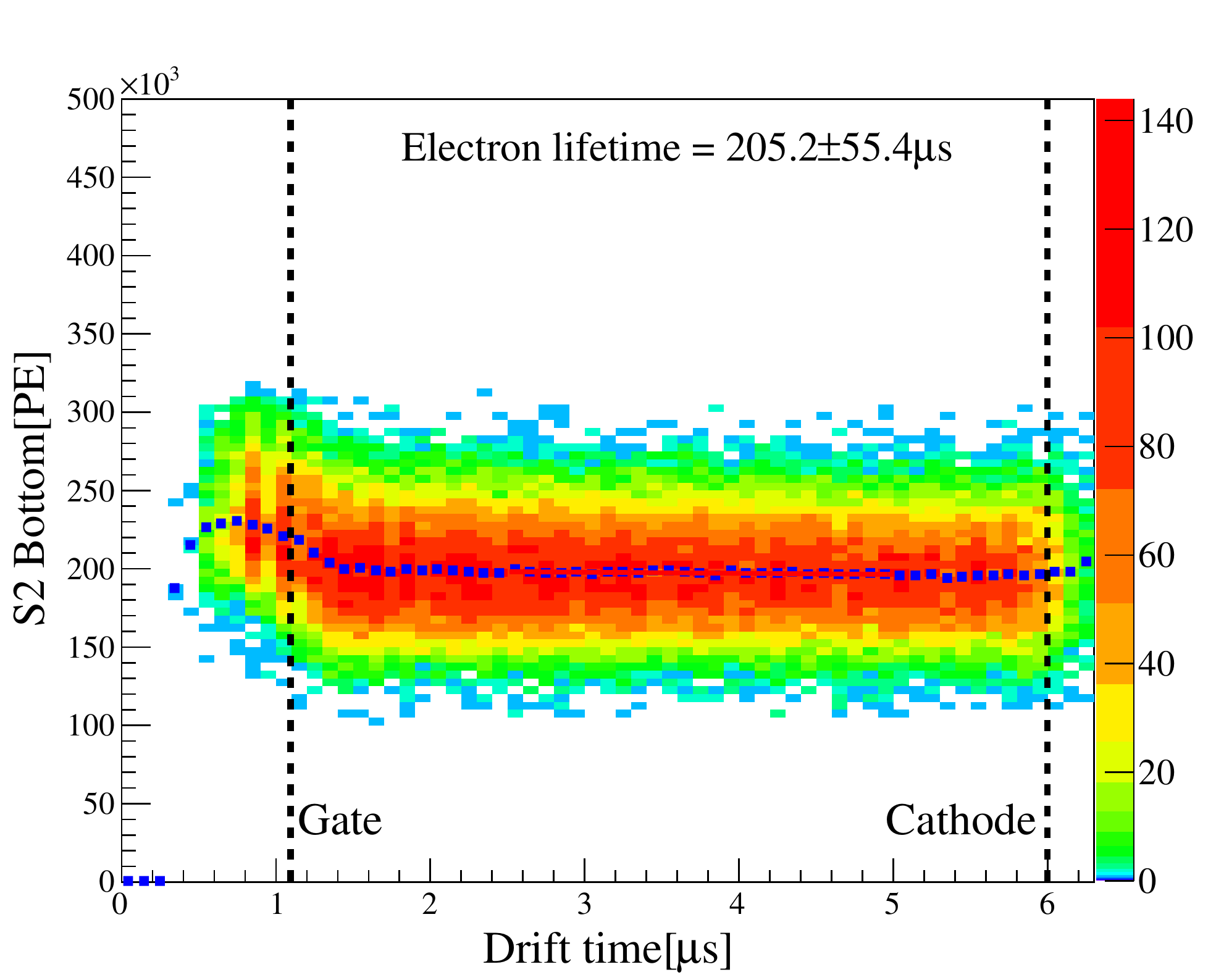}
 \end{minipage}
\hfill
  \begin{minipage}{0.48\textwidth}
   \centering
   \includegraphics[height=5.5cm, width=7.5cm]{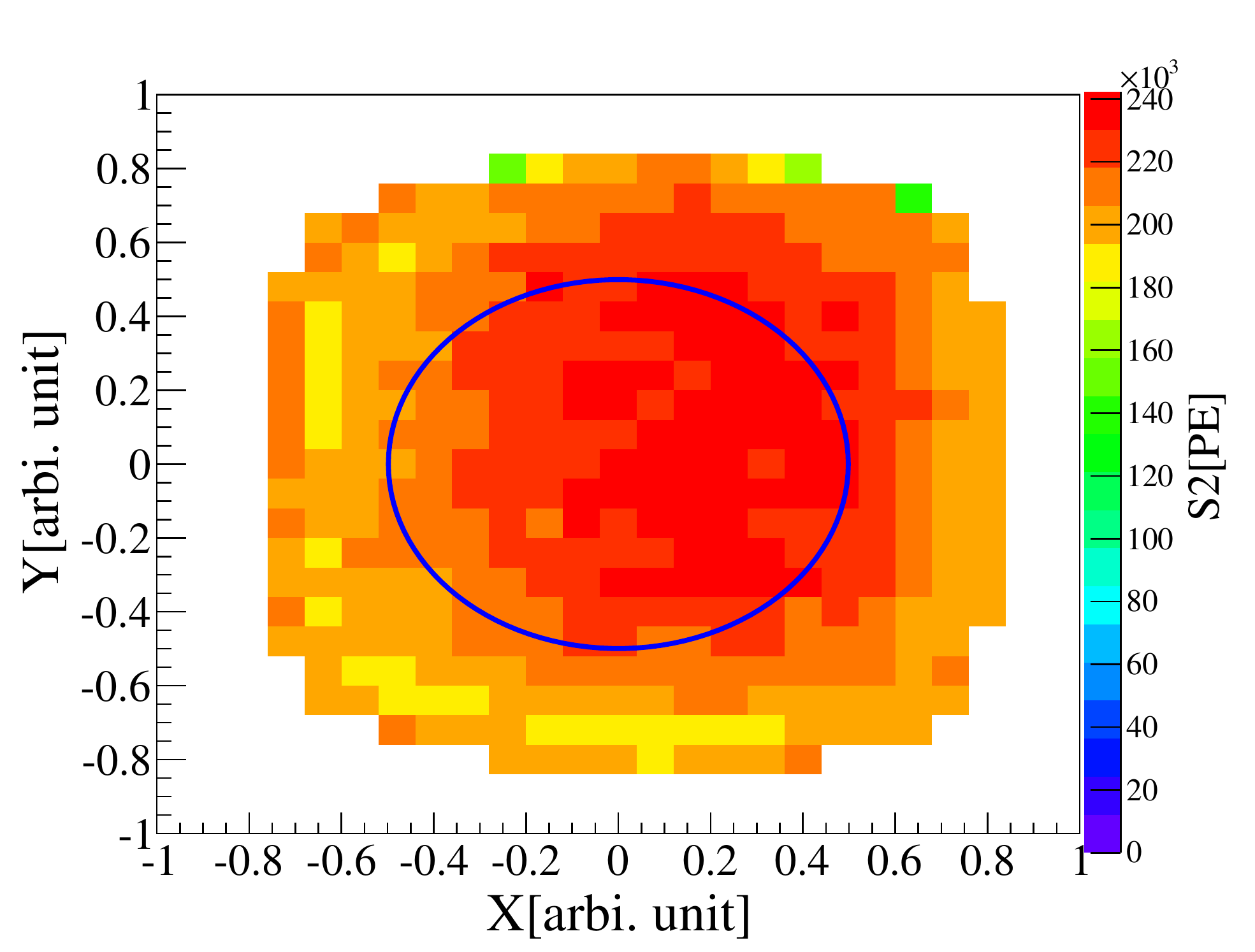}
  \end{minipage}
 \caption{\small (Left) The position dependence of S2 signals from 164\,keV gamma rays at 1kV/cm  drift field.The maximum attenuation of S2 signals for events at the bottom versus those at the top of the sensitive volume is 2.4\%. Blue dots indicate the median S2 values of each drift time slice. An exponential fit was applied for drift time ranging from 2.5\,$\mu s$ to 5\,$\mu s$. Dashed lines indicate the positions of gate mesh and cathode. (Right) X-Y dependence of S2 signals from 164\,keV gamma rays at 1\,kV/cm drift field. The blue solid line indicates the fiducial volume selection.}
 \label{fig:S2posdep}
\end{figure}

 In Fig.~\ref{fig:S2posdep} the position dependence of the S2 signals from the 164\,keV gamma rays in the volume os shown. Unlike for the S1 signal, the S2's position dependence is due to three factors. The first is the dependence on the drift time due to attachment of drifting electrons to impurities in liquid xenon. The second is the S2 production in the gas gap, where a non-uniform gap will result in a non-uniform S2 production. The third is the S2 light collection's position dependence. The S2 dependence on drift time is shown in Fig.~\ref{fig:S2posdep} (left), where an exponential fit of $S2(t_d) = S2(0) \exp(-t_d/\tau)$ results in a measured electron lifetime ($\tau$) of 205\,$\mu$s, improved from 84\,$\mu$s after four days of purification at a flow rate of at least 5\,SLPM. The maximum variation of S2 signals in the entire 5\,$\mu$s drift region is 2.4\%. The right panel of Fig.~\ref{fig:S2posdep} indicates a larger S2 value in the center than near the edge, which is due to a combined effect of non-
uniform S2 light collection efficiency and uneven gas gap for S2 production. The data are corrected for the S2 position dependence due to the non-uniform gas gap and the light collection. The S2 variance (RMS/mean) in the entire region is 6.5\%. After correcting the position dependence using the map in Fig.~\ref{fig:S2posdep}, the S2 variance decreased to 3.2\%.


\subsection{ Energy reconstruction}

\begin{figure}[htp]
 \centering

  \includegraphics[height=7cm, width=14cm]{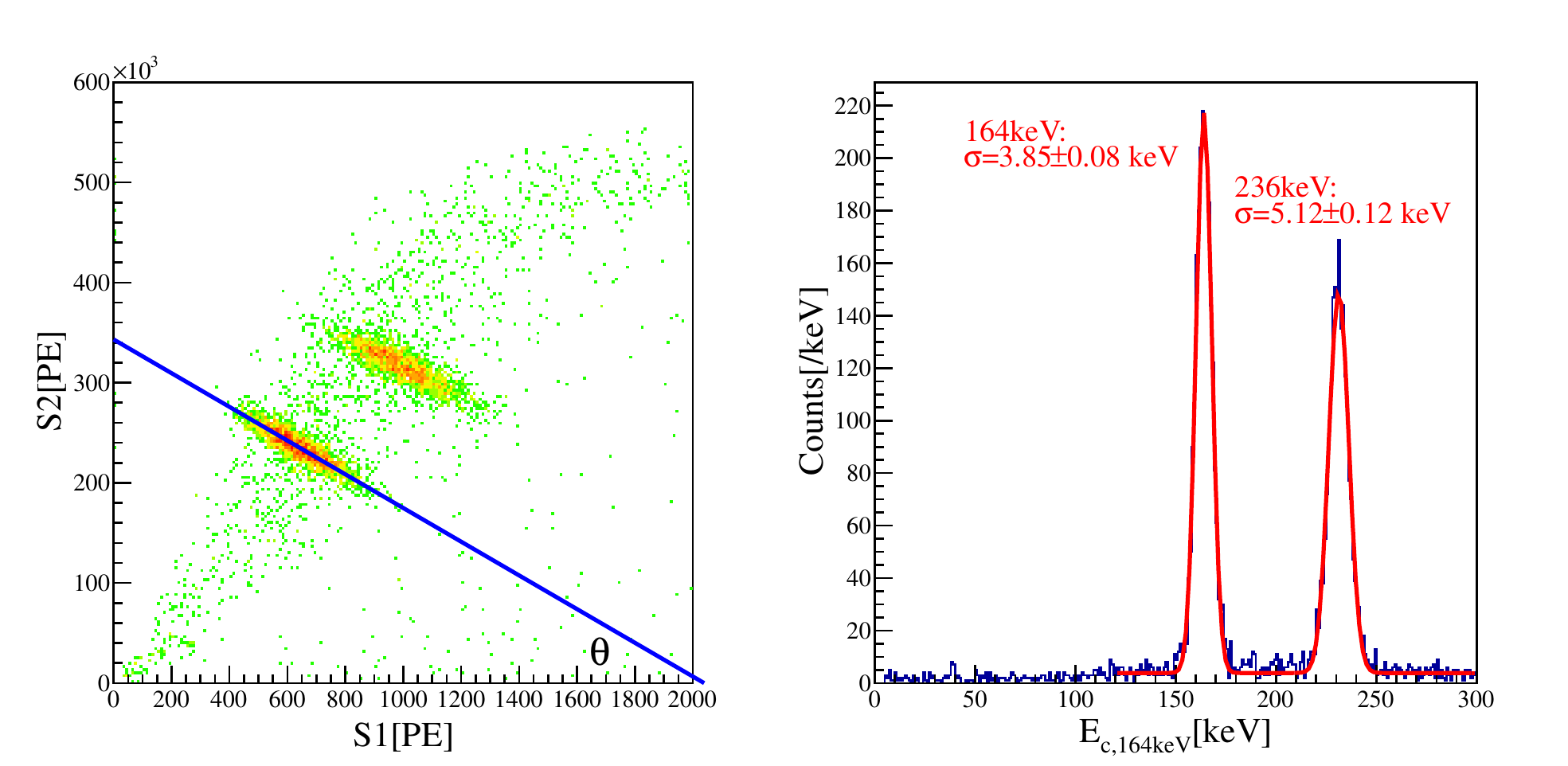}

 \caption{\small (Left) Correlation of the S1 and S2 signals from activated xenon at 2\,kV/cm drift field. A 2-D Gaussian fit was applied to the S2-S1 profile and the correlation angle $\theta$ was obtained. (Right) Combined energy spectrum showing the two activated xenon lines at 164\,keV and 236\,keV. The combined scale was defined based on 164\,keV best fit. A fit using two independent Gaussian functions plus a flat background gives resolutions ($\sigma/E$) of 2.35\% and 2.17\% for the two lines, respectively. The reconstructed energy of 236\,keV under 164\,keV gamma's best fit scale is 231\,keV. The best resolution for 236\,keV under its best fit scale is 2.10\%.}
 \label{fig:S1S2space}
\end{figure}

Because activated xenon atoms are diffused uniformly in liquid xenon, uniform event distribution over the whole sensitive volume was expected. However, events near the edge of the sensitive volume also include background events. To exclude background events, which originate from mostly the enviromental gamma rays and the radioactivity in the detector materials, we select single-scatter events in the center of the TPC, indicated by the lines in Figs.~\ref{fig:S1posdep} and~\ref{fig:S2posdep} (corresponding to a fiducial mass of about 3.2\,grams). The fiducial mass is 4\% of the total sensitive mass of 77\,grams. Additionally, 10\% of events with small width and 30\% of events with large width were rejected to ensure good shapes of S2 pulses. This width cut improves energy resolution by less than 0.1\% for activated xenon lines and by 0.4\% for 662\,keV gamma rays. Fig.~\ref{fig:S1S2space} shows two populations of events in the S2-S1 correlation space after 
the 
data selections described above.

Both S1 and S2 can be used individually to construct the energy scale. However it is better to combine these signals for energy reconstruction\,\cite{Aprile:PRB07, EXO:2013}. It was observed that projecting the events along the S1-S2 correlation axis produces the best energy resolution for calibrated gamma line energies (Fig.~\ref{fig:S1S2space}). Thus the energy scale $E_{c,\langle E \rangle}$, usually referred to as the combined energy scale, is determined by combining the two signals as
\begin{equation}
 E_{c, \langle E \rangle} = \frac{S2+S1\cdot \tan{\theta}}{\langle S2 \rangle + \langle S1 \rangle \cdot \tan{\theta}} \cdot \langle   E \rangle ,
 \label{equ:CES}
\end{equation}
where $\langle E \rangle$ is the energy of the total absorption events; $\langle S1 \rangle$ and $\langle S2 \rangle$ are the mean S1 and S2 values, obtained through a 2-D Gaussian fit; and $\theta$ represents the anti-correlation angle of the S2-S1 profile which is shown in the left plot of Fig.~\ref{fig:S1S2space}.


Fig.~\ref{fig:S1S2space} shows an example of a combined energy spectrum from an activated xenon run at 2\,kV/cm. Energy resolutions of 2.35\% and 2.10\% are obtained for the 164\,keV and 236\,keV lines in the central volume. For comparison, the energy resolutions for these two lines in the entire sensitive volume are 4.4\% and 3.4\%, respectively. This is due to a larger position-dependence of correction and more background contamination events near the edge.

\begin{figure}[htp]
 \centering
  \includegraphics[height=7cm, width=14cm]{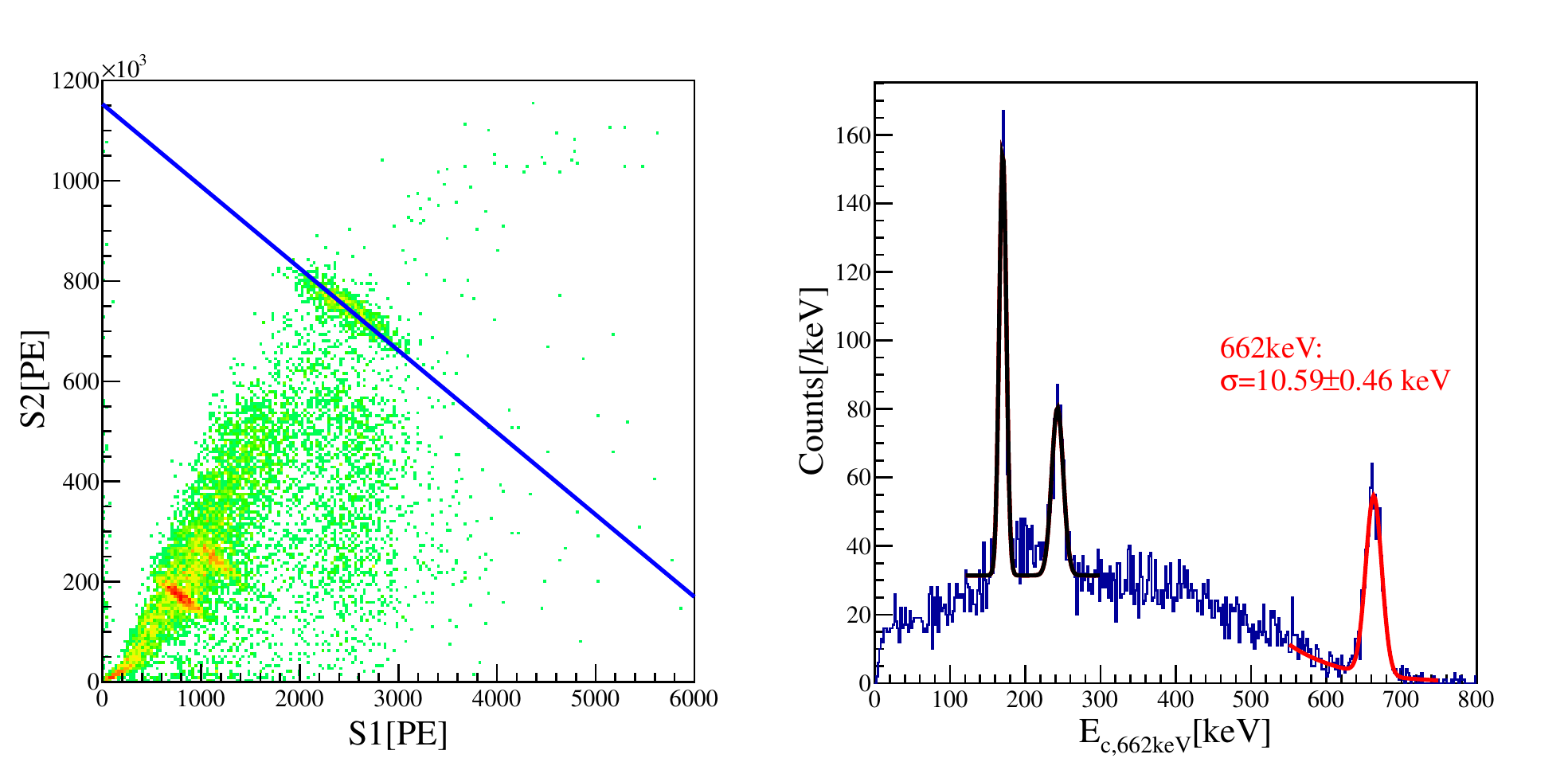}

 \caption{\small (Left) Correlation of S1 and S2 signals for a $^{137}$Cs run with remaining activated xenon at a drift field of 0.5\,kV/cm.  (Right) Combined energy spectrum for the same run. A fit using a Gaussian plus an exponential background gives a resolution($\sigma/E$) of 1.60\% for the 662\,keV peak. The reconstructed energies for the two activated xenon lines at 164\,keV and 236\,keV are 170 and 243\,keV, with Gaussian sigmas of 4.36 and 6.70\,keV, respectively. They deviate slightly from the actual energies due to the energy dependence of the combined energy scale.}
 \label{fig:cs137res}
\end{figure}

Fig.~\ref{fig:cs137res} shows the combined energy spectrum from a $^{137}$Cs run at 0.5\,kV/cm with remaining activity from activated xenon. An energy resolution of $1.60\%$ was achieved. For comparison, the resolution for the 662\,keV peak in the entire volume is 2.42\%. The resolutions for the two activated xenon lines (164\,keV and 236\,keV), with energy reconstruction based on the best-resolution scale for the 662\,keV gamma rays, are slightly worse than their best resolutions for 500\,V/cm drift field shown in Fig.~\ref{fig:worldres} (left).  
These variations in resolution are the result of the energy dependence of the combined energy scale.

\subsection{\label{subsec:Resolution} Energy resolution and drift field dependence}

\begin{figure}[htp]
 \begin{minipage}{0.48\textwidth}
  \centering
  \includegraphics[width=7cm, height=6cm]{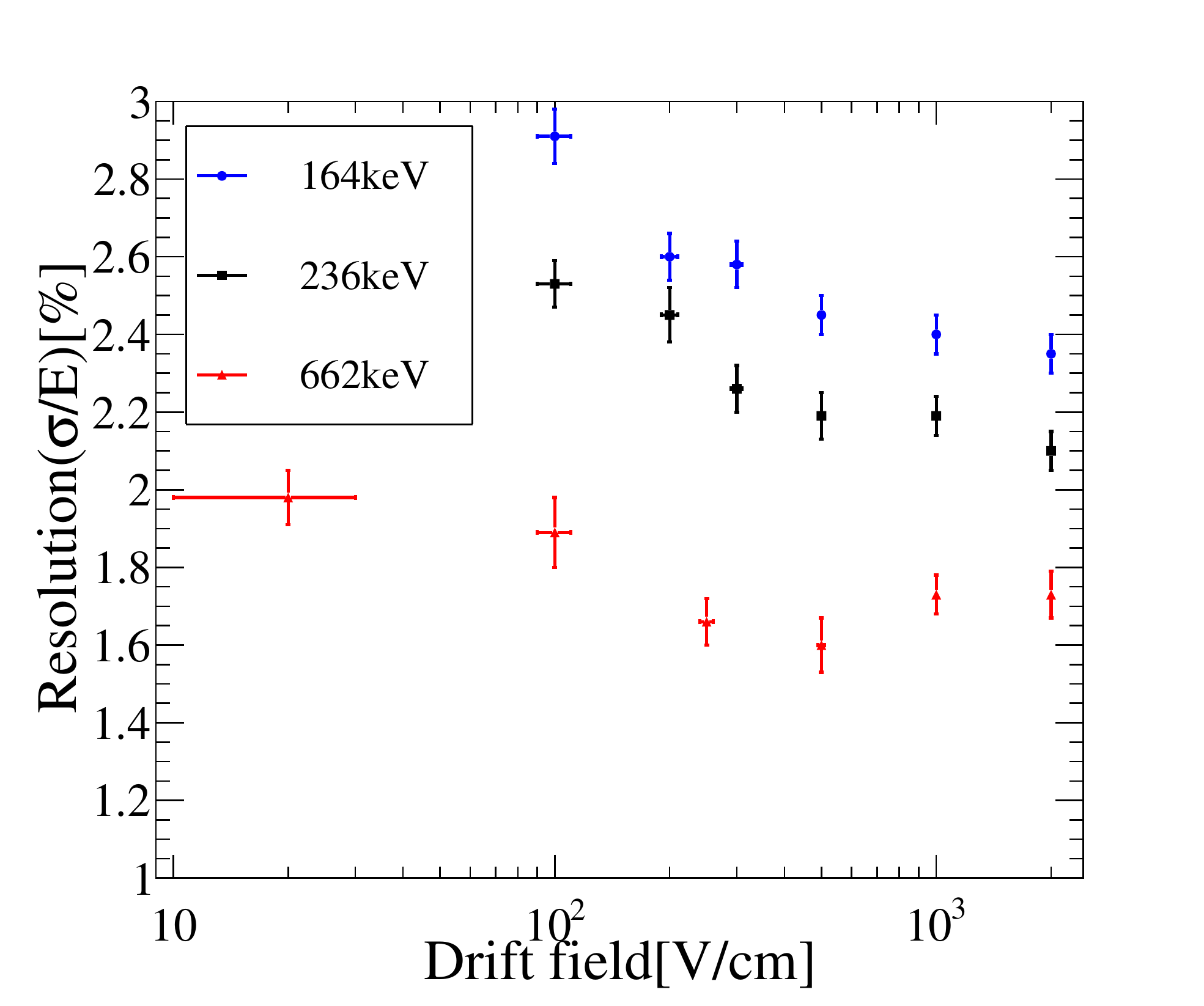}
 \end{minipage}
\hfill
 \begin{minipage}{0.48\textwidth}
  \centering
  \includegraphics[width=8cm, height=6cm]{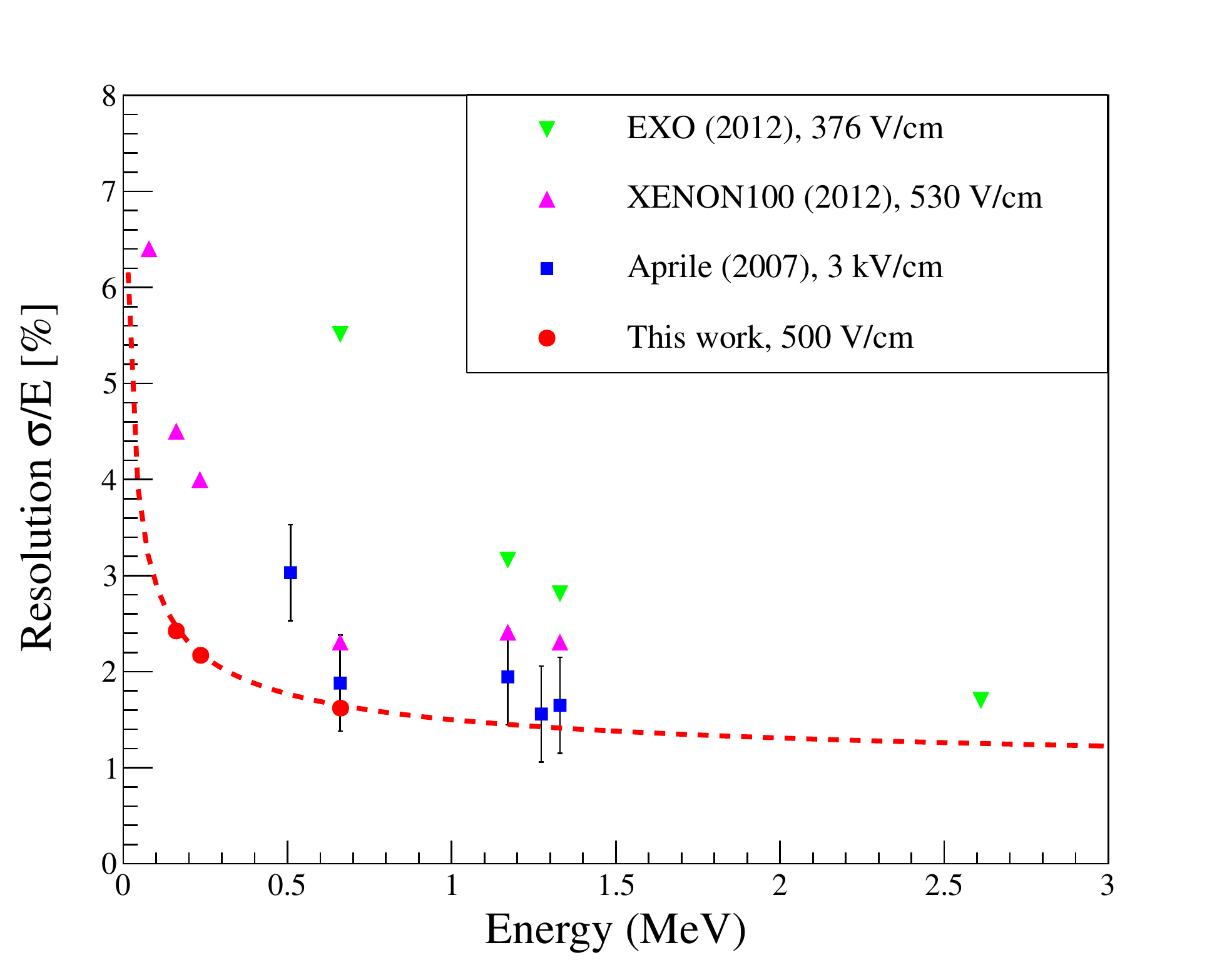}
 \end{minipage}
 \caption{\small (Left) Energy resolution measured for 164, 236, 662\,keV gamma rays at different drift fields. (Right) Energy resolutions in liquid xenon measured by EXO\,\cite{EXO:2013}, XENON100\,\cite{XENON100_tech}, Aprile et al.\,\cite{Aprile:PRB07} and this work. The dashed line indicates a function described by $\sigma/E = (0.65/\sqrt{E} + 0.83$) \%.}
 \label{fig:worldres}
\end{figure}

In order to study the energy resolution as a function of the drift field, we varied the field between 100\,V/cm and 2\,kV/cm. The results is shown in Fig~\ref{fig:worldres}. For drift fields between 100\,V/cm and 2\,kV/cm, better than 2\% resolution can be obtained for 662\,keV gamma rays. We observed a slight improvement from 100\,V/cm to 500\,V/cm, which is due to the increased number of drifted electrons that survive from recombination as the drift field increases. After that, no further improvement was observed because the improvement caused by the increased number of drift electrons was canceled by the reduction in primary scintillation light. For 662\,keV gamma rays, the resolution becomes slightly worse at a higher drift field. This is mainly due to saturation of S2 signals. Additional optimization of the PMT gain and electronic chain is needed to investigate whether the resolution can be further improved. 
According to Ref.\,\cite{ChenDanliThesis}, a ratio of electric field above and below the gate electrode larger than 2.2 is needed for a 100\% electron transmission efficiency. For a given gate voltage (4\,kV) and the gas gap in the setup, a 100\% transmission efficiency can be achieved with a drift field less than 2.4$\pm$0.1\,kV/cm (2.8$\pm$0.2\,kV/cm) for activated xenon ($^{137}$Cs) measurements. For drift field above 3\,kV/cm, the electron transmission drops and the resolution obtained gets worse.
In Fig.~\ref{fig:worldres}, we compare the energy resolution 
for gamma rays of various energies for some of the liquid xenon detectors in the world. The 1.60\% energy resolution for 662\,keV gamma rays obtained at 0.5\,kV/cm drift field in the current study represents the best resolution achieved so far in liquid xenon.  


\section{ Conclusion}
We report a high resolution detection of gamma rays in a dual phase xenon time projection chamber. Based on a simple position reconstruction algorithm using the S2 signals from four R8520 PMTs on the top, mesh wires separated by 2\,mm are clearly visible in central region within a radius of about 15\,mm. This suggests that the X-Y position resolution in the center of the detector can be improved to be better than 0.5\,mm by using a finer mesh. With such a good position resolution, we can correct the position dependence of the signals by using activated xenon lines uniformly distributed in the target volume. 

The 1.60\% ($\sigma/E$) energy resolution for 662\,keV gamma rays is obtained by combining the scintillation and ionization signals from events in central detector volume of 3.2\,grams, for a drift field of 0.5\,kV/cm. Based on the resolutions obtained from three gamma lines, we expect to achieve 1.5\% resolution for 1\,MeV gamma rays and 1.2\% resolution at the Q-value (2458\,keV) of $^{136}$Xe neutrinoless double beta decay. Such an energy resolution makes it possible to search for $^{136}$Xe neutrinoless double beta decay simultaneously in a dual phase xenon detector for dark matter searches, with optimized PMT response for large S2 signals and the capability of correcting signals based on their precisely reconstructed positions. The low field operation with excellent position and energy resolution for gamma rays at hundreds keV region also provides interesting applications of the dual phase XeTPC in gamma ray imaging.

\acknowledgments

We would like to thank Prof. Xuehao Shen for the calibration sources. We would also like to thank Prof. Katsushi Arisaka for valuable suggestions and Fei Gao, Scott Stephenson for useful discussions.

The work is supported by the Ministry of Science and Technology of China (Grant No.: 2010CB833005), National Science Foundation of China (Grant No.: 11055003, 11175117 and 11375114), and Science and Technology Commission of Shanghai Municipality (Grant No.:  11PJ 1405300 and 11DZ2260700).

\end{document}